\documentclass[aps,prl,twocolumn,floatfix,superscriptaddress]{revtex4-1}
\pdfoutput=1
\usepackage{graphicx}
\usepackage{amsmath,amsfonts,amssymb}
\usepackage[colorlinks=true,linkcolor=blue,citecolor=blue]{hyperref}
\usepackage{siunitx}
\usepackage{color,soul}

\begin{document}
\title{In-situ strain tuning in hBN-encapsulated graphene electronic devices}

\author{Lujun Wang}
\email{lujun.wang@unibas.ch}
\affiliation{Department of Physics, University of Basel, Klingelbergstrasse 82, CH-4056 Basel, Switzerland}
\affiliation{Swiss Nanoscience Institute, University of Basel, Klingelbergstrasse 82, CH-4056 Basel, Switzerland}

\author{Simon Zihlmann}
\affiliation{Department of Physics, University of Basel, Klingelbergstrasse 82, CH-4056 Basel, Switzerland}

\author{Andreas Baumgartner}
\affiliation{Department of Physics, University of Basel, Klingelbergstrasse 82, CH-4056 Basel, Switzerland}
\affiliation{Swiss Nanoscience Institute, University of Basel, Klingelbergstrasse 82, CH-4056 Basel, Switzerland}

\author{Jan Overbeck}
\affiliation{Department of Physics, University of Basel, Klingelbergstrasse 82, CH-4056 Basel, Switzerland}
\affiliation{Swiss Nanoscience Institute, University of Basel, Klingelbergstrasse 82, CH-4056 Basel, Switzerland}
\affiliation{Empa, Swiss Federal Laboratories for Materials Science and Technology, Überlandstrasse 129, CH-8600, Dübendorf, Switzerland}

\author{Kenji Watanabe}
\affiliation{National Institute for Material Science, 1-1 Namiki, Tsukuba, 305-0044, Japan}

\author{Takashi Taniguchi}
\affiliation{National Institute for Material Science, 1-1 Namiki, Tsukuba, 305-0044, Japan}

\author{P\'eter Makk}
\email{peter.makk@unibas.ch}
\affiliation{Department of Physics, University of Basel, Klingelbergstrasse 82, CH-4056 Basel, Switzerland}
\affiliation{Department of Physics, Budapest University of Technology and Economics and Nanoelectronics Momentum Research Group of the Hungarian Academy of Sciences, Budafoki ut 8, 1111 Budapest, Hungary}

\author{Christian Sch\"onenberger}
\affiliation{Department of Physics, University of Basel, Klingelbergstrasse 82, CH-4056 Basel, Switzerland}
\affiliation{Swiss Nanoscience Institute, University of Basel, Klingelbergstrasse 82, CH-4056 Basel, Switzerland}

\begin{abstract}
Using a simple setup to bend a flexible substrate, we demonstrate deterministic and reproducible in-situ strain tuning of graphene electronic devices. Central to this method is the full hBN encapsulation of graphene, which preserves the exceptional quality of pristine graphene for transport experiments. In addition, the on-substrate approach allows one to exploit strain effects in the full range of possible sample geometries and at the same time guarantees that changes in the gate capacitance remain negligible during the deformation process. We use Raman spectroscopy to spatially map the strain magnitude in devices with two different geometries and demonstrate the possibility to engineer a strain gradient, which is relevant for accessing the valley degree of freedom with pseudo-magnetic fields. Comparing the transport characteristics of a suspended device with those of an on-substrate device, we demonstrate that our new approach does not suffer from the ambiguities encountered in suspended devices.
\end{abstract}

\maketitle

The large mechanical strength of two-dimensional (2D) crystals allows one to modify their optical and electronic properties by externally induced strain fields \cite{Roldan2015}. Graphene, one of the key examples of 2D materials, is of particular interest because of its peculiar electronic properties \cite{CastroNeto2009}. A series of intriguing effects were predicted for strained graphene, such as the appearance of a scalar potential \cite{Guinea2010}, pseudomagnetic fields \cite{Guinea2009,Guinea2010,Zhu2015}, valley filtering \cite{Milovanovic2016,Yesilyurt2016} or superconductivity \cite{Uchoa2013}. Different methods have been introduced to generate strain in graphene. One common approach is based on suspended graphene, where strain is induced by using different microactuators \cite{PerezGarza2014,Downs2016,Goldsche2018,Colangelo2018} or by simply bending a flexible substrate \cite{Guan2017}. In other approaches, graphene is not suspended and strain can be generated by bending a flexible substrate \cite{Mohiuddin2009}, by using highly stressed metallic pads \cite{Shioya2015}, or by placing graphene on periodic structures \cite{Jiang2017,Liu2018,Zhang2018a}. However, several challenges that need to be overcome simultaneously hampered the progress of these platforms for studying strain effects in transport experiments. First, complex fabrication usually significantly degrades the graphene quality and hinders the observation of the strain effects. In addition, the device is often limited to very basic structures, without the possibility of local gating or multi-terminal devices. Second, mechanical deformations often result in changes in the gate capacitance that cannot be easily distinguished from the actual strain effects. The third challenge is that the strain should be in-situ tunable and non-hysteretic to disentangle strain effects from other effects.

\begin{figure}[htb]
	\centering
	\includegraphics[]{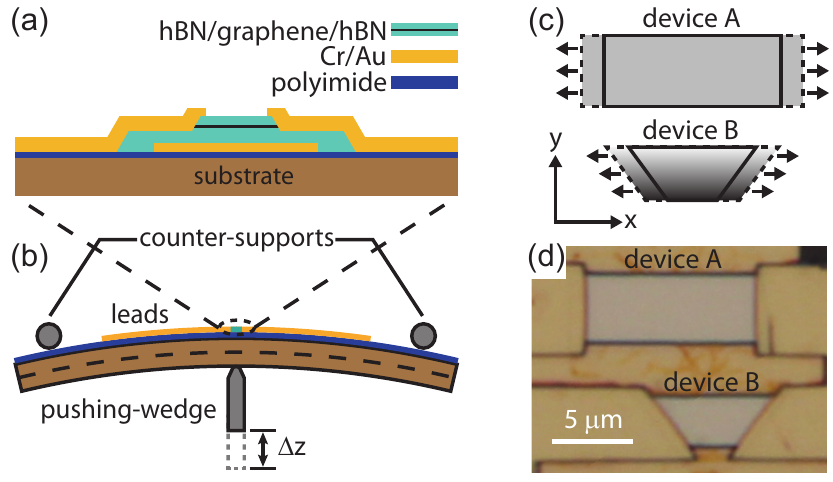}
	\caption{\textbf{(a)} Schematic cross section of our device and \textbf{(b)} of the three-point bending setup. The bending is described by the displacement of the pushing-wedge, $\Delta$z. \textbf{(c)} Illustration of the mechanism for different strain fields. The solid lines represent devices without strain while the dashed lines stand for strained devices. The arrows indicate the elongation of the device along the contacts. The magnitude of strain is shown in grayscale with black corresponding to high strain. \textbf{(d)} Micrograph of two typical devices with different geometries.}
	\label{fig:fig1}
\end{figure}

\begin{figure*}[htb]
	\centering
	\includegraphics[width=17.78cm]{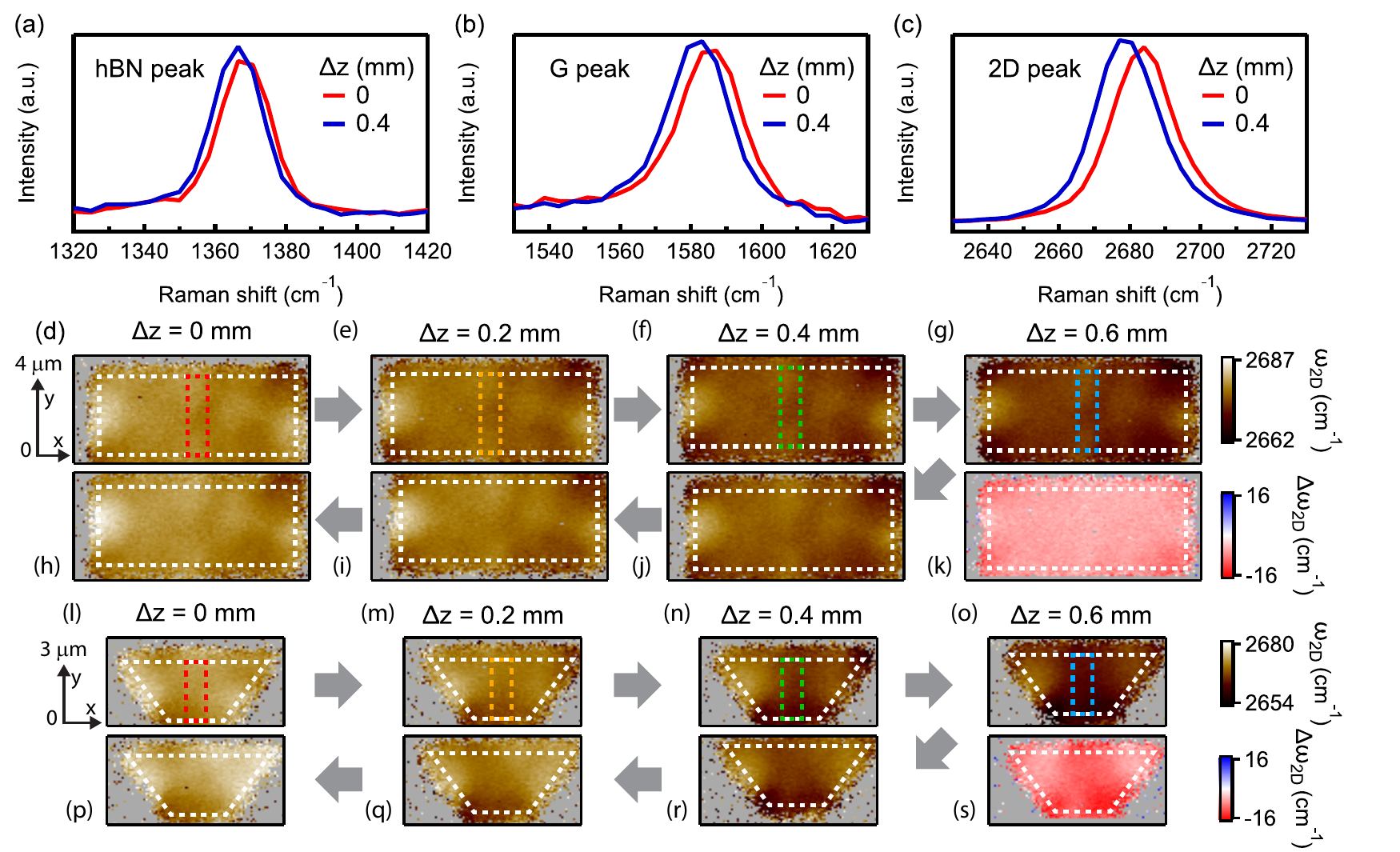}
	\caption{\textbf{(a-c)} Typical Raman spectra of an encapsulated device for two different $\Delta$z values, zoomed in to the hBN peak, the graphene G peak and 2D peak, respectively. \textbf{(d-j)} Spatially resolved Raman maps of $\omega_{2D}$ for device A at different $\Delta$z values. The white dashed lines outline the device boundaries and the colored dashed boxes mark the positions of the profiles shown in Figure \ref{fig:fig3}. The gray arrows show the sequence of the measurements. \textbf{(k)} Spatially resolved Raman maps of $\Delta\omega_{2D}$ for device A, obtained by subtracting map (d) from map (g). \textbf{(l-r)} Spatially resolved Raman maps of $\omega_{2D}$ for device B at different $\Delta$z values. \textbf{(s)} Spatially resolved Raman maps of $\Delta\omega_{2D}$ for device B, obtained by subtracting map (l) from map (o).}
	\label{fig:fig2}
\end{figure*}
 
Here we report a straining method that meets all the above requirements. Instead of suspending the graphene, we encapsulate the graphene with hexagonal boron-nitride (hBN) and directly strain the van der Waals (vdW) heterostructure on-substrate. This approach preserves the exceptional quality of pristine graphene, which is a significant advantage for transport experiments. We use Raman spectroscopy to demonstrate that our method is versatile and that it allows one to engineer various strain fields, such as strain gradients, which are important for the generation of pseudo-magnetic fields \cite{Guinea2010}. We also show that the edge contacts work reliably and can sustain a strain up to $\sim 1$\%. In the first low-temperature electron transport measurements, we demonstrate that our method solves the problem of an artificial gating effect due to the bending-induced change in the graphene-to-gate distance, which is present in suspended graphene devices. Therefore, our approach of on-substrate encapsulated graphene offers an ideal platform for studying strain effects in transport experiments.

As shown schematically in Figure \ref{fig:fig1}a, we fabricate hBN-encapsulated graphene devices with edge contacts \cite{Wang2013} and an electrical bottom gate on a polyimide coated flexible substrate. Bending this substrate in a three-point bending setup, as illustrated in Figure \ref{fig:fig1}b, generates a strain field in the graphene. The deformation of the substrate is determined by the displacement $\Delta$z of the pushing-wedge relative to the mounting position. The hBN encapsulation ensures the high quality of the graphene, and the metallic contacts and the gate allow us to perform transport experiments while tuning the strain in-situ. The metallic contacts are essential for generating strain in a vdW heterostructure. In the Supporting Information Figure S2, we present a comparison between devices with and without contacts. The result shows that strain cannot be induced by substrate bending in devices without contacts. Based on the assumption that the graphene sheet is pulled uniaxially by the contacts during the bending of the substrate, we designed devices with two different geometries in order to obtain different strain fields. This is illustrated in Fig. \ref{fig:fig1}c, where the rectangle (device A) is expected to result in a homogeneous strain field, while the trapezoid (device B) should exhibit a strain gradient along the y-axis, i.e. perpendicular to the straining axis. An optical micrograph of two fabricated devices is shown in Fig. \ref{fig:fig1}d.

\begin{figure*}[htb]
	\centering
	\includegraphics[]{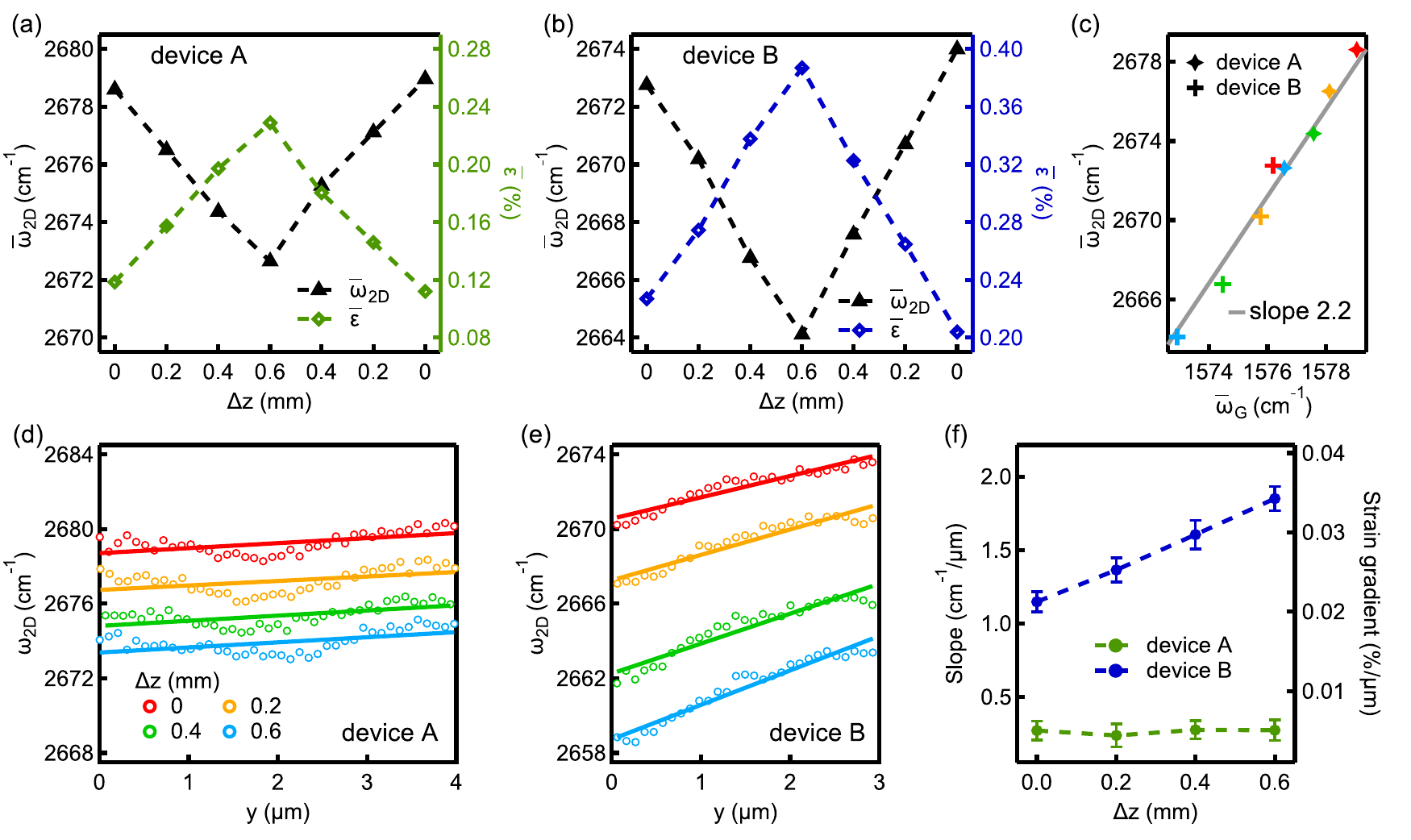}
	\caption{\textbf{(a,b)} $\bar{\omega}_{2D}$ and the corresponding strain values plotted as a function of $\Delta$z for devices A and B, respectively. \textbf{(c)} $\bar{\omega}_{2D}$ versus $\bar{\omega}_{G}$ at different $\Delta$z for both devices, the grey line has a slope of 2.2. \textbf{(d,e)} Profiles at the center of the maps indicated in Figure \ref{fig:fig2} for $\Delta$z = \SI{0}{\milli\meter} (red), 0.2mm (orange), 0.4mm (green), 0.6mm (blue). Open circles are data, solid lines are linear fit. \textbf{(f)} Slopes of the profiles plotted as a function of $\Delta$z. The slopes are extracted from linear fitting in \textbf{(d)} and \textbf{(e)}. The error bars are the fitting errors. The corresponding strain gradient is shown on the right axis. }
	\label{fig:fig3}
\end{figure*}

To characterize the strain fields for different displacements $\Delta$z, we perform Raman spectroscopy directly on a 3-point bending setup at room temperature. Previous studies reported characteristic redshifts in the Raman peaks of graphene \cite{Mohiuddin2009,Yoon2011,Frank2011}, which we now use to quantify the local strain generated in these two devices. Typical Raman spectra of the hBN peak and the graphene G and 2D peaks are shown in Figures \ref{fig:fig2}a-c, respectively, for two different $\Delta$z values. The bending of the substrate results in a redshift of all peaks, with the most prominent effect on the graphene 2D peak. For small strain values the 2D peak can be fitted by a single Lorentzian with center frequency $\omega_{2D}$ \cite{Mohiuddin2009}. In the following, we use spatially resolved Raman spectroscopy to map the strain field based on the redshift of the graphene 2D peak. The same analysis for the hBN Raman peak is presented in the Supporting Information Figure S4, which shows that both hBN and graphene are strained similarly in this method. 

We first focus on the investigation of the rectangular device A. In Figures \ref{fig:fig2}d-j, $\omega_{2D}$ is plotted as a function of position for device A, for a series of increasing $\Delta$z from \SI{0}{\milli\meter} to \SI{0.6}{\milli\meter} (straining) and then decreasing back to \SI{0}{\milli\meter} (relaxing), as indicated by the gray arrows. With increasing $\Delta$z, $\omega_{2D}$ shifts to lower values at all positions on the map, consistent with increasing strain everywhere in the graphene sheet. When $\Delta$z is decreased back to \SI{0}{\milli\meter}, $\omega_{2D}$ reverts back to the initial values. The mean value $\bar{\omega}_{2D}$ averaged over the whole device is plotted as a function of $\Delta$z in Figure \ref{fig:fig3}a, where $\Delta$z is first increased from \SI{0}{\milli\meter} to \SI{0.6}{\milli\meter} then decreased back to \SI{0}{\milli\meter}. The very symmetric V-shape reveals a linear dependence and a good reproducibility of the strain tuning in the graphene device and exhibits no significant hysteresis. The corresponding average strain values ($\bar{\varepsilon}$) are shown on the right axis, which are calculated using $\partial \omega_{2D}/\partial \varepsilon = \SI{-54}{\per\centi\meter}/\%$ \cite{Mohr2009} with $\omega_{2D} \approx (\num{2685} \pm {4.4})\SI{}{\per\centi\meter}$ for unstrained graphene. We note that the former value is not known very accurately and our choice is among the intermediate reported values \cite{Mohiuddin2009,Mohr2009,Huang2010,Yoon2011,Frank2011,Polyzos2015,Colangelo2018}. The latter value is obtained as the average over 10 Raman spectra measured at different positions on three different hBN/graphene/hBN stacks before fabrication. The strain values at $\Delta$z = \SI{0}{\milli\meter} are not 0, probably due to the intrinsic strain accumulated in the device after fabrication. We obtain an average strain of up to $0.23$\% and a maximum strain near the contacts of $0.3$\% for device A at $\Delta$z = \SI{0.6}{\milli\meter}. Figure \ref{fig:fig3}c shows the plot of $\bar{\omega}_{2D}$ versus $\bar{\omega}_{G}$ (center frequency of the graphene G peak) of both devices for different $\Delta$z values. The data points fall on a line of slope 2.2, which confirms strain as the origin of the redshift of the Raman peaks \cite{Lee2012,Neumann2015,Mueller2017,Goldsche2018}.


We note that there is a small inhomogeneity in the map of $\omega_{2D}$ for device A in Figure \ref{fig:fig2}d which does not change significantly with external straining. We analyze the data in the scatter plot of $\omega_{2D}$ versus $\omega_{G}$ in the Supporting Information Figure S3, which shows that the inhomogeneity in the maps of $\omega_{2D}$ originates mostly from the strain variation over the large device area \cite{Lee2012,Neumann2015,Froehlicher2015,Mueller2017}. The homogeneity of the externally induced strain field can be seen directly in figure \ref{fig:fig2}k, which shows a map of the {\it change} in the Raman shift, $\Delta\omega_{2D}$, between the map at $\Delta$z = \SI{0.6}{\milli\meter} (Figure \ref{fig:fig2}g) and the map without external straining (Figure \ref{fig:fig2}d). For this rectangular geometry, the externally induced strain is fairly homogeneous in the bulk with a vanishing strain gradient, which matches quite well the expected strain fields for device A as shown in Figure \ref{fig:fig1}c. The detailed strain pattern for this geometry from finite element method (FEM) simulations is shown in Supporting Information Figure S5.

We now turn to the investigation of strain and strain gradients in the trapezoidal device B. The spatially resolved maps of $\omega_{2D}$ for device B are plotted in Figure \ref{fig:fig2}l-r for the same series of $\Delta$z as above for sample A. Also this device shows a tunable average strain controlled by $\Delta$z. The device averaged $\bar{\omega}_{2D}$ and the corresponding extracted strain values are plotted in Figure \ref{fig:fig3}b as a function of $\Delta$z. At identical $\Delta$z value, the average strain for device B is larger than that for device A due to the smaller size of device B, but shows similar V-shape, i.e. a linear, non-hysteretic dependence on $\Delta$z. These findings can also be seen directly in the Raman maps. We obtain an average strain of up to $0.38$\% and maximum values at the lower sample edge of $0.52$\% for device B at $\Delta$z = \SI{0.6}{\milli\meter}. The existence of a strain gradient is visible in Figure \ref{fig:fig2}m-o. At the shorter (bottom) edge of the device, $\omega_{2D}$ shows a stronger shift than that at the longer (top) edge, which matches the predicted strain pattern for a trapezoidal geometry, as illustrated in Figure \ref{fig:fig1}c (see also the FEM simulations in Supporting Information Figure S5). Figure \ref{fig:fig2}s shows the difference between the Raman signals at large bending (Figure \ref{fig:fig2}o) and no bending (Figure \ref{fig:fig2}l). 


To demonstrate the deterministic generation of a strain gradient in more detail, we plot $\omega_{2D}$ for both devices in Figure \ref{fig:fig3}d \& e as a function of the position along the y-axis in the center of the device area, averaged over $\SI{1}{\micro\meter}$ in the x-direction, as indicated by the colored dashed boxes in Figures \ref{fig:fig2}d-g \& l-o. For both devices we find a clear increase in the average strain (overall shift of the curves) and an essentially unchanged background variation with increasing $\Delta$z. In addition, for the trapezoidal geometry (device B) we find a clear linear increase in the strain when moving from the longer to the shorter sample edge.

\begin{figure}[htb]
	\centering
	\includegraphics[]{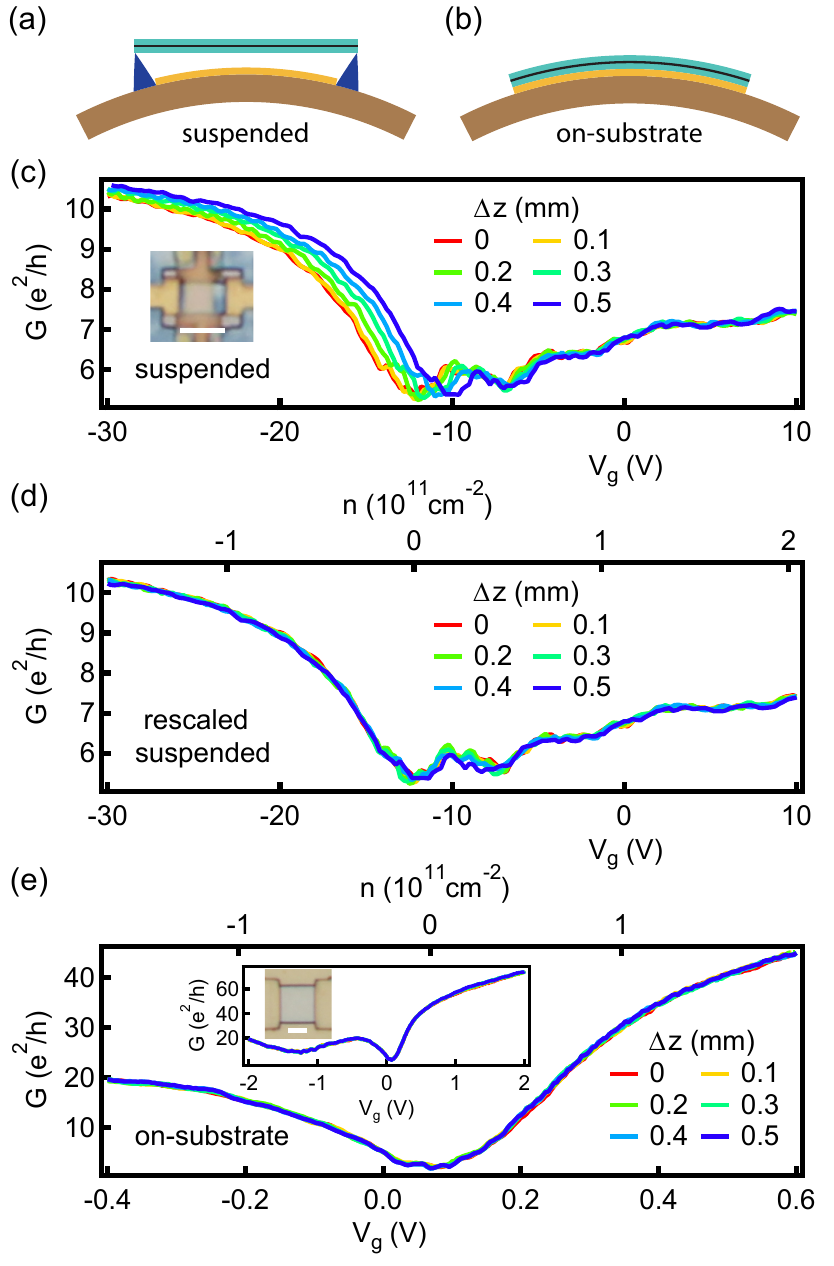}
	\caption{\textbf{(a,b)} Illustration of substrate bending for suspended and on-substrate devices, respectively. The graphene-to-gate distance changes with substrate bending for suspended devices while it remains unchanged for on-substrate devices. \textbf{(c)} Two-terminal differential conductance, $G$,  plotted as a function of gate voltage, $V_g$, for a suspended device for different $\Delta$z values. The inset is the micrograph of the measured device. \textbf{(d)} The same data as in \textbf{(c)} with the curves rescaled in $V_g$ with respect to $V_g$ = 0 by matching the CNP of each curve with that of the curve at $\Delta$z = 0. The corresponding carrier density is shown on the top axis. \textbf{(e)} $G$ of an on-substrate device for a similar charge carrier density range for different $\Delta$z values. The inset shows $G$ on a larger $V_g$ range. The micrograph is the measured device. The scale bars correspond to $\SI{2}{\micro\meter}$.}
	\label{fig:fig4}
\end{figure}

We now take the average slope of these curves as an estimate of the large scale (non-microscopic) strain gradient along the y-axis. For this purpose we plot in Figure \ref{fig:fig3}f the slopes of linear fits to the data in Figure \ref{fig:fig3}d (device A) and Figure \ref{fig:fig3}e (device B) as a function of $\Delta$z, with the right axis showing the corresponding extracted strain gradient. The small non-zero slope for device A stems from the small intrinsic strain variation over the large device area discussed above and stays constant for increasing $\Delta$z. In contrast, for device B we find a linear dependence of the average slope on the $\Delta$z, demonstrating that strain gradients can be generated by modifying the device geometry. The pseudo-magnetic field, $B_{ps}$, depends on the strain gradients, $B_{ps} = \frac{\hbar\beta}{ea}\left [-2\frac{\partial\varepsilon_{xy}}{\partial x}- \frac{\partial(\varepsilon_{xx}-\varepsilon_{yy})}{\partial y} \right ]$ \cite{Guinea2009,Guinea2010}, with $\beta \approx 3.37$ the Gr\"uneisen parameter \cite{Pereira2009a} and $a = \SI{1.42}{\angstrom}$ the interatomic distance. To estimate $B_{ps}$, we use $B_{ps} \approx  \frac{\hbar\beta}{ea}\frac{\partial\varepsilon_{xx}}{\partial y}$ and obtain values on the order of $\SI{10}{mT}$ in the bulk of device B at $\Delta$z = $\SI{0.6}{\milli\meter}$.. We note that much larger strain gradients occur near the corners of the device which might result in pseudo-magnetic fields over $\SI{100}{mT}$ locally. We also point out that larger average strain values can be achieved in this setup (see Supporting Information Figure S6), but with an increased probability of device failure, which happens at a strain on the order of $1\%$.





One of the major advantages of our technique is that the edge contacts not only act as clamps for the mechanical deformation, but allow for simultaneous transport experiments. Here we report first low-temperature ($\SI{4}{\kelvin}$) transport experiments with in-situ strain tuning of a device fabricated in the described fashion and compare the results to similar measurements on a device where the encapsulated graphene is suspended, using the LOR based suspension technique \cite{Tombros2011,Maurand2014}. The two experiments are shown schematically in Figure \ref{fig:fig4}a \& b. For both devices the two-terminal differential conductance, $G$, is measured as a function of the gate voltage, $V_g$, for different $\Delta$z values, see Figures \ref{fig:fig4}c \& e, respectively.

There are significant differences between suspended and on-substrate devices for bending experiments. For the suspended device one can immediately find a systematic change of the curves in gate voltage with increasing $\Delta$z (see Figure \ref{fig:fig4}c), while such an obvious effect is absent for the on-substrate device (see Figure \ref{fig:fig4}e). The effects found in the suspended device can be fully accounted for by the change in the graphene-to-gate distance when bending the substrate, as depicted in Figure \ref{fig:fig4}a. This is illustrated in Figure \ref{fig:fig4}d, where we plot the data of Figure \ref{fig:fig4}c rescaled linearly in gate voltage for each curve with $V_g = \SI{0}{\volt}$ as a fix point. This can be understood in a simple capacitor model in which the charge induced in the graphene are given by $Q = CV_g$, with $C$ the effective capacitance between the graphene layer and the gate. If the capacitance is changed by a factor $\alpha$ to $\alpha C$ due to the substrate bending, the same charge $Q$ is induced at $V_g/\alpha$, which is equivalent to a rescaling in the gate voltage. This scaling factor is extracted for each curve by matching the CNP to that of the curve at $\Delta$z = \SI{0}{\milli\meter} and it is linear in $\Delta$z. After rescaling, all data points fall onto the same curve, see Fig. \ref{fig:fig4}d. This demonstrates that the bending-induced gating effect is dominant for the suspended graphene device, which makes it very difficult to study effects due to actual strain.

 

This effect is absent in the on-substrate devices optimized for strain tuning. For comparison, we performed the same type of measurements also on an on-substrate device, with the results shown in Figure \ref{fig:fig4}e. Since the gate voltage lever arm in this device is much larger than that for the suspended device due to the shorter graphene-to-gate distance, we apply smaller gate voltages to obtain a similar carrier density range as that in the data of the suspended device (see top axes of Figure \ref{fig:fig4}d \& e). The electron mobility of $\SI{\sim100000}{\square\centi\meter\per\volt\per\second}$ is extracted from a linear fit around the CNP, suggesting a high graphene quality in our device. On this gate voltage scale and also on much larger scale (see inset of Figure \ref{fig:fig4}e), all curves with different $\Delta$z values are virtually identical, which demonstrates that there are no bending-induced changes in the carrier density or in the contact resistance in this experimental configuration. The additional conductance minimum at $V_g \approx \SI{1.3}{\volt}$ comes from a double moir\'e superlattice effect in encapsulated graphene \cite{Wang2019}. We point out that on this scale of graphene straining we could not detect significant changes in the conductance.

In conclusion, we have successfully generated tunable and reversible strain fields in encapsulated graphene devices. Since these devices are fabricated on-substrates, i.e. not suspended, it allows us to design a large variety of device geometries. As an example, we use spatially resolved Raman imaging to demonstrate that the edge contact clamping and rectangular geometry result in a fairly homogeneous straining of the graphene. In a second step, we use this design freedom to generate a strain gradient in a trapezoidal geometry. In first transport experiments we then demonstrate another major advantage of on-substrate encapsulated devices, namely that the bending-induced gate capacitance change can be avoided, which is crucial for studying strain effects in transport experiments. This approach is not limited to graphene, but also suitable for studying strain effects in other 2D materials and complex vdW heterostructures, for example in MoS$_2$ \cite{Conley2013,He2013,Zhu2013}. Because our method is simple and intuitive, nonetheless allowing complex device structures, we expect that it will pave the way towards deterministic strain engineering and new approaches to valleytronics.

\subsection{Author contributions}
L.W. fabricated the devices, performed the measurements and did the data analysis. S.Z. initiated the idea of hBN-encapsulation. A.B. did the FEM simulations. P.M and J.O. helped to develop the straining method. P.M., S.Z. and J.O. supported the device fabrication. A.B., P.M. and S.Z. helped with the data analysis. K.W. and T.T. provided high-quality hBN. C.S. initiated and supervised the project. L.W. and A.B. wrote the paper. All authors discussed the results and worked on the manuscript. All data in this publication are available in numerical form at DOI:\href{https://doi.org/10.5281/zenodo.2636531}{10.5281/zenodo.2636531}.

\section*{Acknowledgement}
This work has received funding from the Swiss Nanoscience Institute (SNI), the ERC project TopSupra (787414), the European Union Horizon 2020 research and innovation programme under grant agreement No. 785219 (Graphene Flagship), the Swiss National Science Foundation, the Swiss NCCR QSIT, Topograph, ISpinText FlagERA network and from the OTKA FK-123894 grants. P.M. acknowledges support from the Bolyai Fellowship, the Marie Curie grant and the National Research, Development and Innovation Fund of Hungary within the Quantum Technology National Excellence Program (Project Nr. 2017-1.2.1-NKP-2017-00001). Growth of hexagonal boron nitride crystals was supported by the Elemental Strategy Initiative conducted by the MEXT, Japan and the CREST (JPMJCR15F3), JST. The authors thank Oliver Braun, David Indolese and Peter Rickhaus for fruitful discussions, and Sascha Martin and his team for their technical support.

\bibliography{StrainRaman}

%
%
%

\end{document}


\beginsupplement

\title{Supporting information for \\ In-situ strain tuning in hBN-encapsulated graphene electronic devices}


\author{Lujun Wang}
\email{lujun.wang@unibas.ch}
\affiliation{Department of Physics, University of Basel, Klingelbergstrasse 82, CH-4056 Basel, Switzerland}
\affiliation{Swiss Nanoscience Institute, University of Basel, Klingelbergstrasse 82, CH-4056 Basel, Switzerland}

\author{Simon Zihlmann}
\affiliation{Department of Physics, University of Basel, Klingelbergstrasse 82, CH-4056 Basel, Switzerland}

\author{Andreas Baumgartner}
\affiliation{Department of Physics, University of Basel, Klingelbergstrasse 82, CH-4056 Basel, Switzerland}
\affiliation{Swiss Nanoscience Institute, University of Basel, Klingelbergstrasse 82, CH-4056 Basel, Switzerland}

\author{Jan Overbeck}
\affiliation{Department of Physics, University of Basel, Klingelbergstrasse 82, CH-4056 Basel, Switzerland}
\affiliation{Swiss Nanoscience Institute, University of Basel, Klingelbergstrasse 82, CH-4056 Basel, Switzerland}
\affiliation{Empa, Swiss Federal Laboratories for Materials Science and Technology, Überlandstrasse 129, CH-8600, Dübendorf, Switzerland}

\author{Kenji Watanabe}
\affiliation{National Institute for Material Science, 1-1 Namiki, Tsukuba, 305-0044, Japan}

\author{Takashi Taniguchi}
\affiliation{National Institute for Material Science, 1-1 Namiki, Tsukuba, 305-0044, Japan}

\author{P\'eter Makk}
\email{peter.makk@unibas.ch}
\affiliation{Department of Physics, University of Basel, Klingelbergstrasse 82, CH-4056 Basel, Switzerland}
\affiliation{Department of Physics, Budapest University of Technology and Economics and Nanoelectronics Momentum Research Group of the Hungarian Academy of Sciences, Budafoki ut 8, 1111 Budapest, Hungary}

\author{Christian Sch\"onenberger}
\affiliation{Department of Physics, University of Basel, Klingelbergstrasse 82, CH-4056 Basel, Switzerland}
\affiliation{Swiss Nanoscience Institute, University of Basel, Klingelbergstrasse 82, CH-4056 Basel, Switzerland}

\maketitle


\section{Fabrication}\label{sec:fab}
\begin{figure}[H]
    \centering
      \includegraphics[width=\columnwidth]{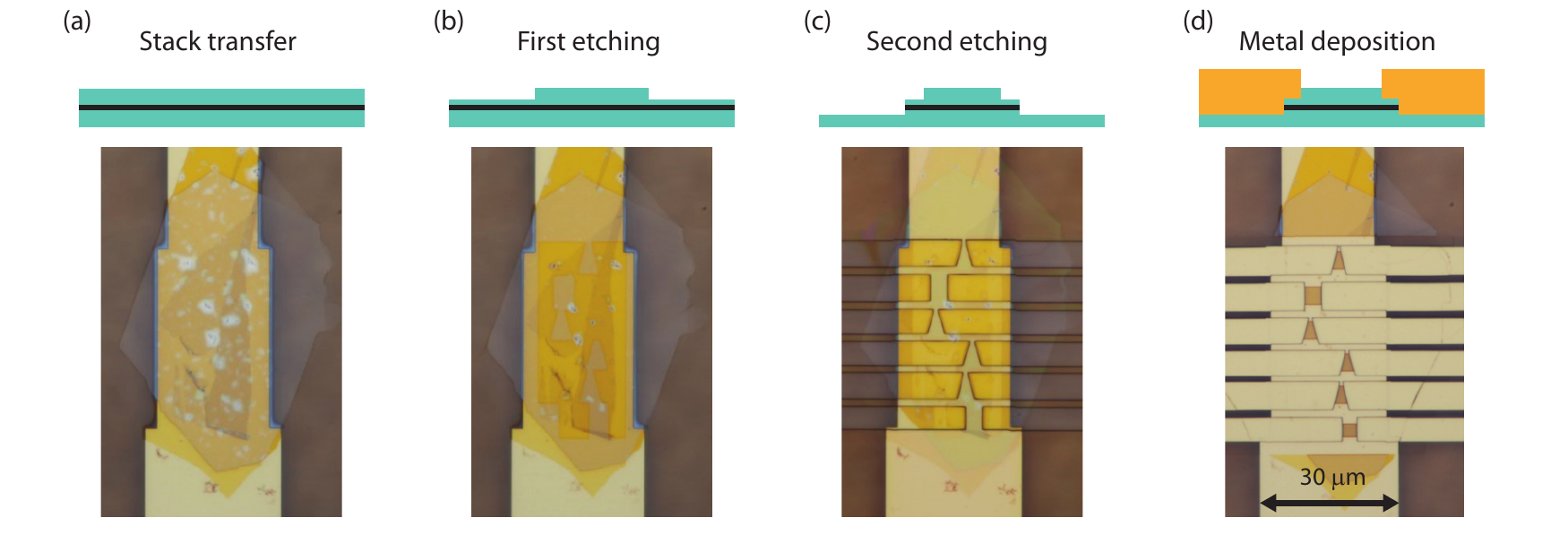}
    \caption{Schematics and corresponding micrographs of different fabrication steps. The overlaps near the edge contacts are designed for mechanical reinforcement.}
    \label{fig:fabrication}
\end{figure}

We use a polished phosphor bronze plate with dimension of $\SI{24}{mm}\times\SI{9.5}{mm}\times\SI{0.3}{mm}$ as the flexible substrate. A layer of $\SI{\sim5}{\micro\meter}$ polyimide is spin coated on the substrate for electrical isolation. The metallic gate is fabricated in the center of the substrate using standard e-beam lithography (EBL) and metalization (Cr/Au, $\SI{5}{nm}/\SI{75}{nm}$). The hBN/graphene/hBN stack is prepared with the pick-up technique \cite{Wang2013,Zomer2014} and then placed on the gate, as shown in Figure \ref{fig:fabrication}a. We first use one EBL step to etch the top hBN down to $\SI{\sim10}{nm}$ above the graphene (see Figure \ref{fig:fabrication}b). In the second EBL step we open the mask $\SI{\sim400}{nm}$ more so that it overlaps with the unetched top hBN and etch another $\SI{\sim15}{nm}$ to cut the graphene for making the edge contacts (Cr/Au, $\SI{10}{nm}/\SI{110}{nm}$), as depicted in Figure \ref{fig:fabrication}c-d.

\section{Raman Measurements}\label{sec:Raman}
The Raman spectra were acquired using a commercially available confocal Raman system WiTec alpha300 with the excitation laser wavelength of $\SI{532}{\nano\meter}$ at room temperature. We use an incident laser power of $\SI{1}{\milli\watt}$ with the laser spot size around $\SI{500}{\nano\meter}$. All measurements were performed with linearly polarized light. The grating of the spectrometer is 600 grooves/mm. 

\section{Contact clamping}\label{sec:clamping}
\begin{figure}[H]
    \centering
      \includegraphics[width=\columnwidth]{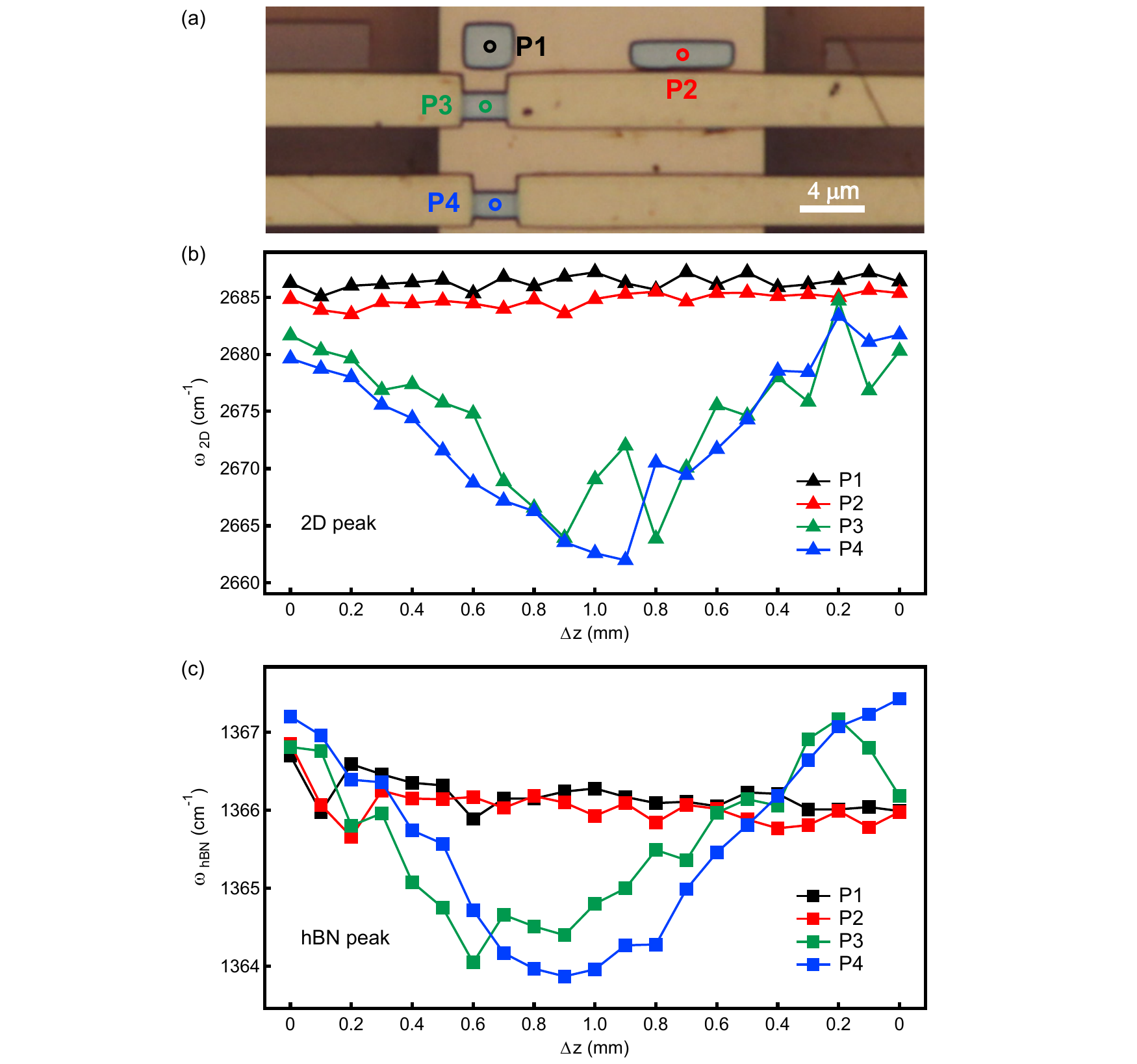}
    \caption{\textbf{(a)} Micrograph of the sample with two devices without contacts (P1, P2) and two devices with contacts (P3, P4). The circles mark the laser spot positions for the Raman measurements. \textbf{(b)} Center frequency $\omega_{2D}$ of the graphene 2D peak plotted as a function of $\Delta$z for all four devices. \textbf{(c)} Center frequency $\omega_{hBN}$ of the hBN peak plotted as a function of $\Delta$z for all four devices.}
    \label{fig:clamp}
\end{figure}

Here we compare the devices with edge contacts to devices without any contacts in Raman measurements. The micrograph of the devices are shown in Figure \ref{fig:clamp}a. During the measurements, the substrate bending, $\Delta$z, is first increased to $\SI{1}{mm}$ and then decreased back to $\SI{0}{mm}$ in steps. In Figure \ref{fig:clamp}b, the $\omega_{2D}$ of device P3 and P4 shows the V-shape behavior, revealing a linear dependence and a good reversibility of the strain tuning in the contacted devices, similar to the devices shown in the main text. The fluctuations on the curves can be attributed to the relative shift of the laser spot between each measurement and the not fully homogeneous strain within the device. In contrast, $\omega_{2D}$ in the devices P1 and P2 do not show any dependence on $\Delta$z, suggesting no strain induced in these devices by bending the substrate. Similar results are observed in the hBN Raman peak, as shown in Figure \ref{fig:clamp}c. These results demonstrate that the contact clamping is essential for generating strain in encapsulated devices.

\section{Scatter plot of $\omega_{2D}$ vs $\omega_{G}$}\label{sec:2D_vs_G}
\begin{figure}[H]
    \centering
      \includegraphics[width=\columnwidth]{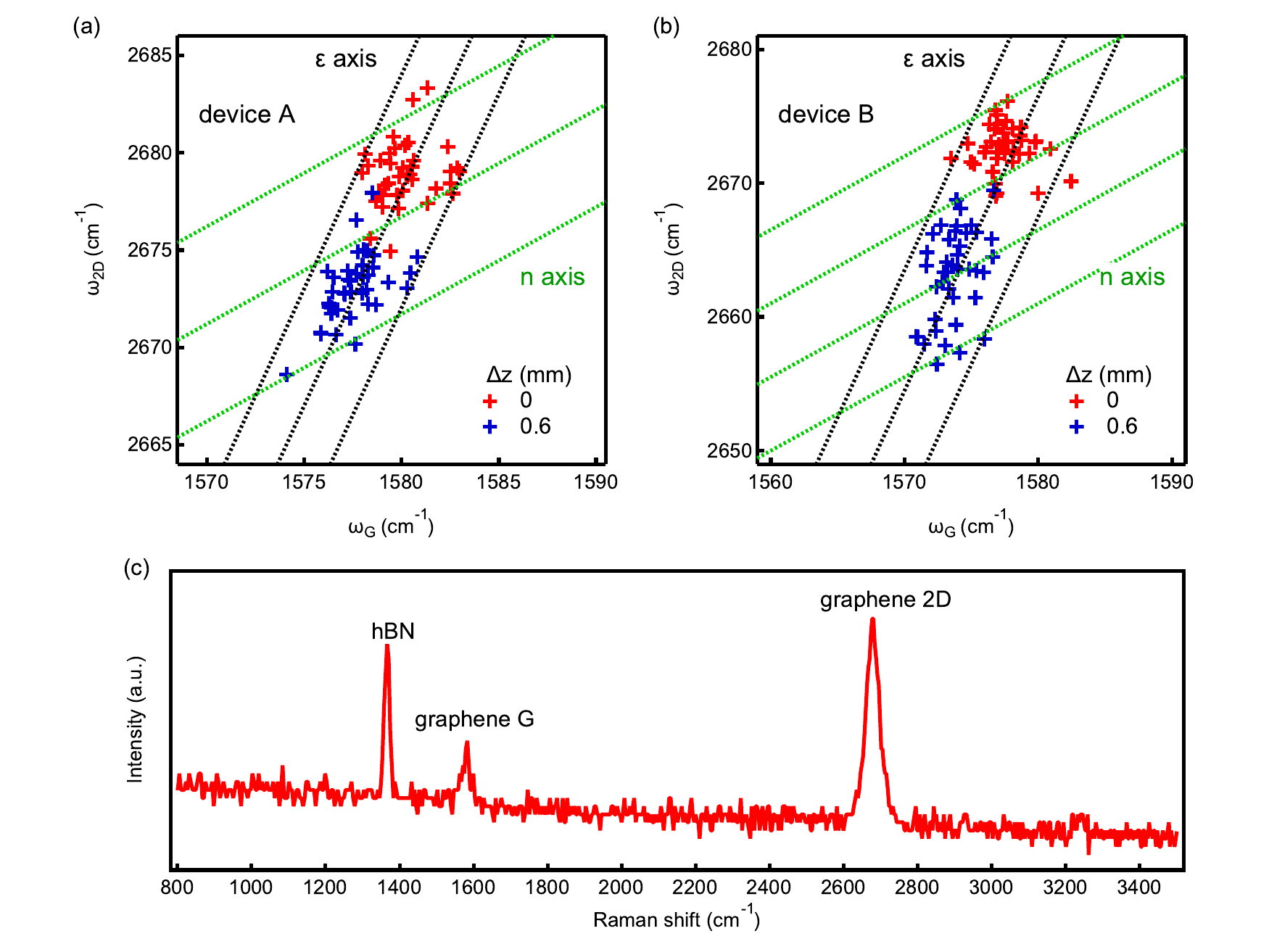}
    \caption{Scatter plot of $\omega_{2D}$ versus $\omega_{G}$ at two different $\Delta$z values for device A \textbf{(a)} and device B \textbf{(b)} in the main text. Data points are taken from 40 different positions homogeneously distributed on each map. The $\Delta\omega_{2D}/\Delta\omega_{G} = 2.2$ lines (black dashed) correspond to the strain axis and the $\Delta\omega_{2D}/\Delta\omega_{G} = 0.55$ lines (green dashed) correspond to the hole doping axis.\textbf{(c)} Typical Raman spectrum from one of the maps.}
    \label{fig:scatter_plot}
\end{figure}  

Here we discuss the inhomogeneity in the maps in Figure 2 of the main text by analyzing the scatter plots of $\omega_{2D}$ versus $\omega_{G}$. Both strain and charge doping can induce a shift of the 2D and G peaks. For strain, the data points are expected to lie along the line with a slope of $\Delta\omega_{2D}/\Delta\omega_{G} \approx 2.2$ \cite{Neumann2015,Lee2012,Mueller2017,Goldsche2018} while the slope is expected to be around 0.55 for hole doping \cite{Froehlicher2015}. Our on-substrate encapsulated devices are predominately hole doped. In Figure \ref{fig:scatter_plot}a, the data points of device A are spread along the strain axis, which corresponds to a strain variation of about $0.1\%$ over the whole device area. With increasing $\Delta$z, the data points shift globally along the strain axis while the spread along the strain axis does not change, which is in good agreement with a homogeneous external strain induced by the substrate bending. There is also a spread of the data points along the doping axis, which would correspond to a doping variation of $\SI{\sim5e12}{\per\square\centi\meter}$ \cite{Froehlicher2015}. However, this doping variation is not realistic, because it is more than one order of magnitude higher than the typical residual doping in our devices. A typical Raman spectrum from one of the maps is shown in Figure \ref{fig:scatter_plot}c, where the graphene G peak is not well resolved due to short integration time during the measurement. Therefore, we attribute the large spread of the data points along the doping axis to the uncertainties in extracting $\omega_{G}$. For device B, in addition to the global shift, the spread of the data points along the strain axis extends with increasing $\Delta$z (see Figure \ref{fig:scatter_plot}b), which is consistent with the externally induced strain gradient from the geometry design. In conclusion, the inhomogeneity in the maps in Figure 2 of the main text is mostly coming from long range strain variation.  

\section{Maps of hBN Raman peak}\label{sec:hBN}
\begin{figure}[H]
    \centering
      \includegraphics[width=\columnwidth]{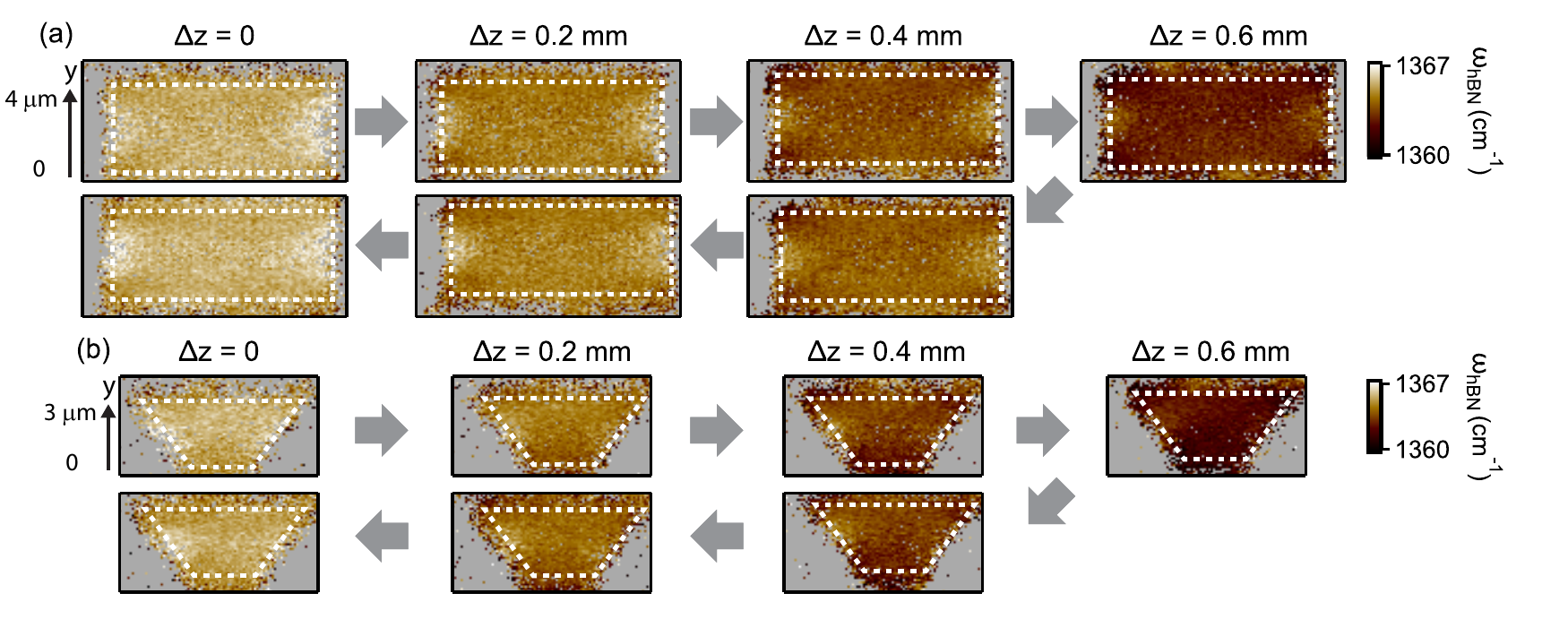}
    \caption{\textbf{(a)} Spatially resolved Raman maps of $\omega_{hBN}$ for device A at different $\Delta$z values. The white dashed lines outline the device boundaries and the gray arrows show the sequence of the measurements. \textbf{(b)} Spatially resolved Raman maps of $\omega_{hBN}$ for device B.}
    \label{fig:CuBJ10_hBN}
\end{figure} 

In this section we plot the center frequency of the hBN Raman peak, $\omega_{hBN}$, as a function of position for device A and device B of the main text. The displacement $\Delta$z is first increased from $\SI{0}{mm}$ to $\SI{0.6}{mm}$ and then decreased back to $\SI{0}{mm}$, as indicated by the gray arrows. The $\omega_{hBN}$ shifts to lower values with increasing $\Delta$z at all positions on the map for both devices, consistent with increasing strain everywhere in the hBN \cite{Androulidakis2018}. The $\omega_{hBN}$ reverts back to the initial values, when $\Delta$z is decreased back to $\SI{0}{mm}$, demonstrating a good reversibility of the strain tuning. In device A, the externally induced strain is fairly homogeneous in the bulk with no strain gradient (see Figure \ref{fig:CuBJ10_hBN}a). In device B, $\omega_{hBN}$ shows a larger red-shift on the bottom edge than on the top edge, demonstrating an externally induced strain gradient (see Figure \ref{fig:CuBJ10_hBN}b). These results are consistent with those extracted from the graphene 2D peak in the main text. 

\section{FEM simulations of strain}\label{sec:FEM}
\begin{figure}[H]
    \centering
      \includegraphics[width=\columnwidth]{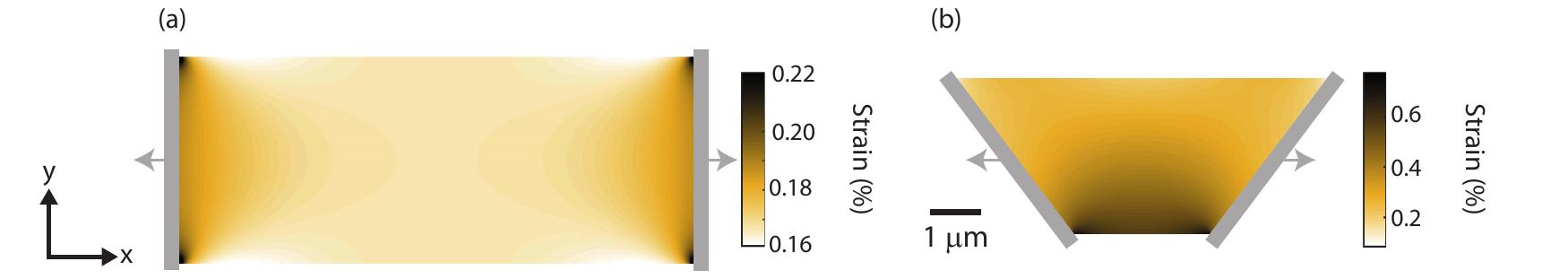}
    \caption{FEM simulations of strain for device A \textbf{(a)} and device B \textbf{(b)}in the main text.}
    \label{fig:FEM}
\end{figure}

Here we perform the strain simulations for the two geometries given in the main text. We use Matlab’s finite element method (FEM) modeling toolbox to solve the 2D partial differential equations in the plane strain mode to estimate the strain distribution in the limit of a continuous, homogeneous and isotropic medium. The elasticity parameters we use are a 2D Young’s modulus of $\SI{340}{N/m}$, a Poisson’s ratio of 0.165 \cite{Lee2008,Blakslee1970}, and no volume forces. We set the boundary conditions such that at the edge contacts (gray beams in Figure \ref{fig:FEM}) are displaced by $\SI{10}{nm}$ in the direction of the gray arrows, and no normal forces act on all other boundaries. We plot the hydrostatic strain ($\varepsilon_{xx}+\varepsilon_{yy}$), the quantity expected to result in the Raman 2D shift \cite{Mueller2017}. These settings reproduce the measured strain distributions quite well, which suggests a strong clamping mediated by the edge contacts. We note, however, that quantitatively, $\varepsilon_{xx}$ alone resembles the Raman data more closely, which we tentatively attribute to the Raman laser being linearly polarized along the uniaxial strain direction (x-direction) \cite{Yoon2008,Huang2010,Yoon2011}. 

\section{Contact failure}\label{sec:failure}
\begin{figure}[H]
    \centering
      \includegraphics[width=\columnwidth]{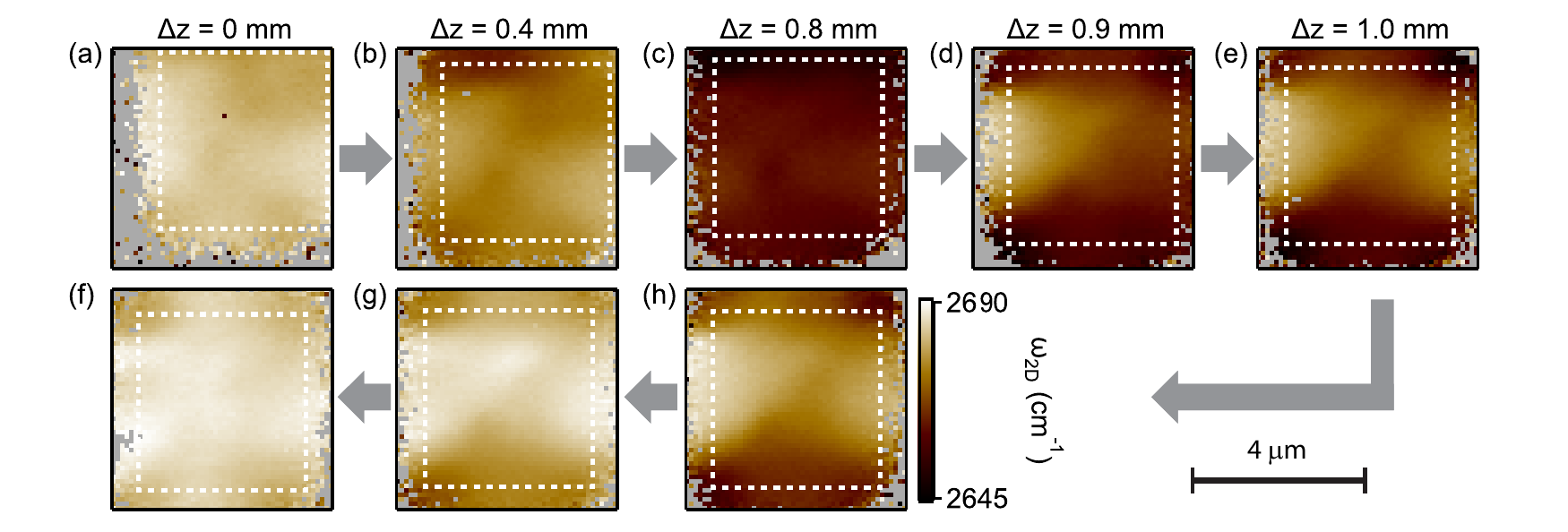}
    \caption{Spatially resolved Raman maps of $\omega_{2D}$ for a rectangle device at different $\Delta$z values. The white dashed lines outline the device boundaries and the gray arrows show the sequence of the measurements.}
    \label{fig:contac_failure}
\end{figure}

Here we show a typical straining sequence leading to contact failure of a device at large bending. In Figure \ref{fig:contac_failure}a-c, $\omega_{2D}$ shifts to lower values with increasing $\Delta$z for the whole device area, as expected for an externally induced strain in the graphene. After a critical amount of bending, here around $\SI{0.9}{mm}$, the $\omega_{2D}$ does not red-shift with increasing bending, but instead blue shifts. This indicates a failure of the mechanical contact to the metal. This process is gradual as one can see in Figure \ref{fig:contac_failure}d\&e. When the contact failure happens, the electrical contact resistance increases dramatically. The average strain achieved in this device is about 0.7\% before the contact failure. 

\section{Extended discussion of the experiments with the suspended encapsulated device}\label{sec:suspended}
\begin{figure}[H]
    \centering
      \includegraphics[width=\columnwidth]{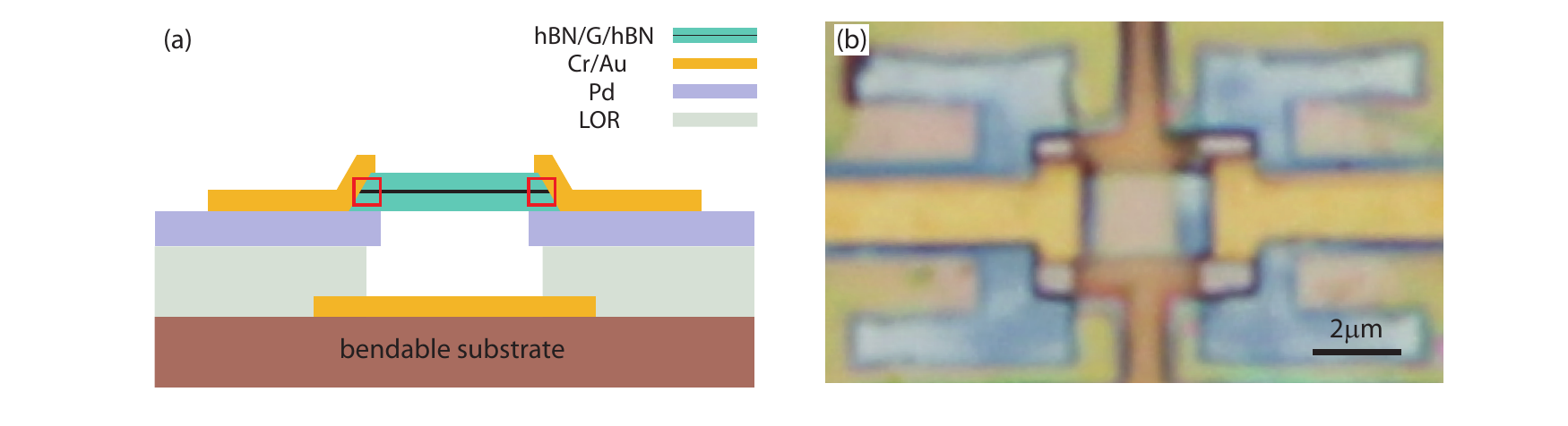}
    \caption{Schematics and micrograph of the suspended encapsulated device. The red boxes mark the screened region which cannot be tuned by the bottom gate.}
    \label{fig:suspended}
\end{figure}

The schematics and the micrograph of the suspended device discussed in the main text is shown in Figure \ref{fig:suspended}. The Pd support is used to protect the LOR from UV exposure during the etching of the stack. In Figure 4c of the main text, we note that the conductance is generally lower on the electron side for the suspended device compared to that of the on-substrate device. We attribute this to a region screened by the Pd, as marked by red boxes in Figure \ref{fig:suspended}a. These screened regions cannot be tuned by the bottom gate and can form an additional p-n junction when the device is tuned to the electron side, which then suppresses the total conductance of the device.

Usually, in our on-substrate encapsulated devices, the graphene is hole doped at zero gate voltage. In the suspended encapsulated devices, the graphene is electron doped (see Figure 4c in main text), which is usually the case in suspended graphene devices fabricated with the LOR based suspension technique \cite{Maurand2014}. The electron doping might be attributed to LOR residuals.

\bibliography{StrainRaman}